\documentclass[12pt]{article}
\newcommand{\be}{\begin{equation}}
\newcommand{\ee}{\end{equation}}
\newcommand{\bea}{\begin{eqnarray}}
\newcommand{\eea}{\end{eqnarray}}
\newcommand{\bt}{\begin{tabular}}
\newcommand{\et}{\end{tabular}}
\newcommand{\ba}{\begin{array}}
\newcommand{\ea}{\end{array}}
\newcommand{\ov}{\overline}
\newcommand{\bvec}{\mathbf}

\begin{document}
\setcounter{page}{0}
\thispagestyle{empty}
\baselineskip=20pt
%---------------------------------------------------------------------------

\hfill{
\begin{tabular}{l}
DSF$-$97/6 \\
INFN$-$NA$-$IV$-$97/6 \\
hep-ph/9705351
\end{tabular}}

\bigskip\bigskip

\begin{center}
\begin{huge}
{\bf Pontecorvo neutrino-antineutrino oscillations: theory and experimental 
limits }
\end{huge}
\end{center}

\vspace{2cm}

\begin{center}
{\Large
Salvatore Esposito 
\footnote{e-mail: sesposito@na.infn.it}
and Nicola Tancredi 
\footnote{e-mail: tancredi@na.infn.it} \\} 
\end{center}

\vspace{0.5truecm}

\normalsize
\begin{center}
{\it
\noindent
Dipartimento di Scienze Fisiche, Universit\`a di Napoli ``Federico 
II''\\
Mostra d'Oltremare Pad. 19, I-80125 Napoli Italy \\
and \\
Istituto Nazionale di Fisica Nucleare, Sezione di Napoli\\
Mostra d'Oltremare Pad. 20, I-80125 Napoli Italy }
\end{center}

\vspace{3truecm}

%%\centerline{{\bf Abstract}}
%%\noindent

\begin{abstract}
We study Pontecorvo neutrino-antineutrino oscillations both in vacuum 
and in matter within a field theoretic approach, showing that this
phenomenon can occur only if neutrinos have a Dirac-Majorana mass 
term. We find that matter effects suppress these oscillations and 
cannot explain the solar neutrino problem. On the contrary, a vacuum 
neutrino-antineutrino oscillations solution to this problem exists. We 
analyze this solution and available data from laboratory 
experiments giving stringent limits on $\nu_e$ and $\nu_{\mu}$ 
Majorana masses.
\end{abstract}

\vspace{1truecm}
\noindent

\newpage

\section{Introduction}

One of the most striking problems in Elementary Particle Physics is that
of neutrino mass. In the Standard Model by Glashow, Weinberg and Salam
\cite{GWS} neutrino are massless, but there is no physical explanation for this
fact. On the contrary, in the framework of Grand Unified Theories massive
neutrinos appear to be more natural \cite{Buccella} and, moreover, when
quantum gravity effects are taken into account, a non vanishing neutrino mass
term naturally arises \cite{Barbierietal}.

Furthermore, massive neutrinos are also required in Cosmology and Astrophisics
to explain the large scale structure and hydrogen ionization in the universe
(see \cite{Smirnov} and ref. therein) and peculiar velocities of pulsar
\cite{Segre}. From the experimental point of view, indications of a 
non-vanishing neutrino mass came from the so called ``solar neutrino problem"
\cite{SNP} and atmospheric neutrino flux anomaly \cite{atmo}.

Actually, for example, the most simple and natural explanation 
for the solar $\nu_e$ flux deficit is given in term of neutrino flavour 
oscillations in the solar medium \cite{Espo}, according to the MSW mechanism
\cite{MSW}, occurring only if neutrinos are massive.

From a phenomenological point of view, the problem of neutrino mass is 
strictly related to lepton number conservation (both individual
( $L_e$, $L_\mu$, $L_\tau$ ) and total (L) ), and the Dirac or Majorana
nature of neutrinos themselves. In fact, if neutrinos are massless, the 
conservation of all lepton numbers ($L_e$, $L_\mu$, $L_\tau$, L) is allowed,
according to the Glashow-Weinberg-Salam theory, due to the invariance of the 
electroweak lagrangian for un arbitrary (global) phase transformation of the
matter field. Instead, if neutrino have a non-vanishing mass and are of the 
Dirac type (particle states different from the antiparticle ones), so 
that to the e.w. lagrangian the following mass term is added :
\be
{\cal L}^{D}_{m} \; = \; -
\sum_{l,l^{\prime}} 
\ov{\nu}_{l^{\prime}R} \; 
M^{D}_{l^{\prime}l} \;
\nu_{l L} \;
+ \; h.c.
\label{11}
\ee
with $l,l^{\prime} = e,\mu,\tau$ and $M^D$, in general, a complex non diagonal 
3x3 matrix, individual lepton number is no longer conserved 
($L^{D}_{m}$ is not invariant for $\nu_l \rightarrow 
e^{i\alpha_l}\nu_l$), while the total 
lepton number is still conserved ($L^{D}_{m}$ is invariant for 
$\nu_l \rightarrow e^{i\alpha}\nu_l$). 
However, if neutrinos are massive Majorana 
particles (particle states coincide whith antiparticles ones), so that the 
mass term is 
\be
{\cal L}^{M}_{m} \; = \; - \;
\frac{1}{2} \;
\sum_{l,l^{\prime}} 
\ov{\nu}^{c}_{l^{\prime}R} \; 
M^{M}_{l^{\prime}l} \;
\nu_{l L} \;
+ \; h.c.
\label{12}
\ee
with $ M^{M} $, in general, a non symmetric 3x3 matrix, 
no lepton number is conserved ( $L_{m}^{M}$ is not
invariant for every (global) phase transformation). Note that, while the 
term (\ref{12}) does not require additional neutrino states besides 
those present in the Standard Model, the term (\ref{11}) involves right-handed
neutrino states (and related antiparticles) not present in tha S.M. but 
predicted in many GUTs (see for example \cite{Buccella}). 

The phenomenological consequences of eq. (\ref{11}) or 
(\ref{12}) are very intriguing.

For istance, if $M^{D}$ and $M^{M}$ are non diagonal (in analogy to what
happens with the quark mass matrix), neutrino flavour oscillations are 
predicted \cite{GribBil}. But, furthermore, many other processes involving
charged lepton, which violate lepton number conservation, are allowed.
For example, with Dirac neutrinos, the decays 
$\mu \rightarrow e \gamma$, $\mu \rightarrow 3e $, 
$\tau \rightarrow e \pi^{0}$ and the conversion $\mu$-e in presence of nuclei,
such as $\mu^{-} + Ti \rightarrow e^{-} + Ti$, can be realized, while with
Majorana neutrinos neutrinoless double beta decay 
(Z,A) $\rightarrow$ (Z+2,A) + 2$e^{-}$ and reaction 
$\mu^{-}$ + Ti $\rightarrow e^{+}$ + Ca can also occur.

More in general, one can consider a lagrangian mass term whith both Dirac and 
Majorana terms
\be
-{\cal L}^{DM}_{m} \; = \;
\sum_{l,l^{\prime}} 
\ov{\nu}_{l^{\prime}R} 
\; M^{D}_{l^{\prime}l} \;
\nu_{l^{\prime}L} \; + \;
\frac{1}{2} \;
\sum_{l,l^{\prime}} 
\ov{\nu}^{c}_{l^{\prime}R} \; 
M^{1}_{l^{\prime}l} \;
\nu_{l L} \; + \;
\frac{1}{2} \;
\sum_{l,l^{\prime}} 
\ov{\nu}^{c}_{l^{\prime}R} 
\; M^{2}_{l^{\prime}l} \;
\nu_{l R} \; + \; h.c.
\label{13}
\ee
involving $\nu_{L}$ and $\nu^{c}_{R}$, as well as $\nu_{R}$ and $\nu^{c}_{L}$ 
predicted in many GUTs.
This scenario, from a theoretical point of view, allows to give a very small
mass to neutrinos in a very natural way through the so-called ``see-saw"
mechanism \cite{seesaw}. The mass eigenstates coming from (\ref{13})
are in general Majorana states, and the phenomenology previous described 
in the discussion of  eq. (\ref{12}) 
applies in this case as well. However, as we will show in this 
paper, another interesting phenomenon (also violating the total lepton number)
is now possible: neutrino-antineutrino oscillations.
Using field-theoretic methods, in section II we study neutrino-antineutrino
oscillations in vacuum, rederiving the formula for the survival probability
first introduced by Pontecorvo \cite{Pontecorvo}.
In section III we shall extend the results
found to keep into account matter effects for neutrinos propagating
in a medium.
In section IV we analyze the available experimental data and give limits
on the $\nu_{e}$ Majorana mass. Finally the conclusions.\\

\section{Neutrino-antineutrino oscillations in vacuum}

We add the Dirac-Majorana 
mass term (\ref{13}) for neutrinos to the standard electroweak 
lagrangian. Since here we are interested only to neutrino-antineutrino 
oscillations, let us assume that the mass matrices appearing in 
(\ref{13}) are diagonal and concentrate our attention on only one 
flavour at a time. So the propagation of a neutrino in vacuum is 
described by the lagrangian
\be
{\cal L} \; = \; \ov{\nu} \, {\not k} \, \nu \; + \; \ov{\nu^C} \, 
{\not k} \, \nu^C \; - \; \frac{1}{2} \, m_D \, \left( \ov{\nu} \, \nu \; 
+ \; \ov{\nu^C} \, \nu^C \right) \, - \, \frac{1}{2} \, m_M \, \left( 
\ov{\nu} \, \nu^C \; + \; \ov{\nu^C} \, \nu \right)
\label{21}
\ee
where $k_{\mu} = ( \omega ,${\bf k}$)$ is the neutrino 4-momentum and 
$m_D , m_M$ are respectively the Dirac and Majorana mass for the given 
flavour. The neutrino equations of motion are then:
\bea
{\not k} \, \nu \; - \; \frac{1}{2} \, m_D \, \nu \; - \; 
\frac{1}{2} \, m_M \, \nu^C & = & 0  \\
\label{22}
{\not k} \, \nu^C \; - \; \frac{1}{2} \, m_D \, \nu^C \; - \; 
\frac{1}{2} \, m_M \, \nu & = & 0
\label{23}
\eea
equivalent to 
\bea
\left( {\not k} \; - \; m_+ \right) \, \left( \nu \, + \, \nu^C 
\right) & = & 0 \\
\label{24}
\left( {\not k} \; - \; m_- \right) \, \left( - \,\nu \, + \, \nu^C 
\right) & = & 0
\label{25}
\eea
where
\be
m_{\pm} \; = \; \frac{1}{2} \, \left( m_D \; \pm \; m_M \right)
\label{26}
\ee
From (\ref{24}), (\ref{25}) it is easy to recognize that the mass 
eigenstates are Majorana states, $(\nu \, + \, \nu^C)$ and 
$( - \nu \, + \, \nu^C)$, with $C = +1$ and $C = -1$ respectively.

Let us now adopt the chiral Weyl base for the Dirac gamma matrices and
define:
\be
\nu \; = \; \left( \ba{c}
                   \nu_R \\
                   \nu_L    \ea \right) \;\;\;\;\;\;\;\;\;\;\;\;\;\;\;
\nu^C \; = \; \left( \ba{c}
                   \nu^C_L \\
                   \nu^C_R    \ea \right)
\label{27}
\ee
We introduce the helicity eigenstates
\be
\frac{\bvec{\sigma} \cdot \bvec{k}}{k} \, \phi_{\lambda} \; = \; 
\lambda \, \phi_{\lambda}
\label{28}
\ee
with $\lambda = \pm 1$, and write
\bea
\nu_R \; = \; x \, \phi_{\lambda} & \;\;\;\;\;\;\;\;\;\;\;\;\;\;\; &
\nu^C_L \; = \; z \, \phi_{\lambda} \\
\label{29}
\nu_L \; = \; y \, \phi_{\lambda} & \;\;\;\;\;\;\;\;\;\;\;\;\;\;\; &
\nu^C_R \; = \; w \, \phi_{\lambda} 
\label{210}
\eea
the equations of motion (\ref{24}), (\ref{25}) imply:
\bea
\left( \ba{cc}
       \omega \, + \, \lambda k  &  - m_+ \\
       - m_+  &  \omega \, - \, \lambda k
       \ea \right) \left( \ba{c}
                          n_1 \\
                          n_2
                          \ea \right) & = & 0
\label{211} \\
\left( \ba{cc}
       \omega \, + \, \lambda k  &  - m_- \\
       - m_-  &  \omega \, - \, \lambda k
       \ea \right) \left( \ba{c}
                          \widetilde{n_1} \\
                          \widetilde{n_2}
                          \ea \right) & = & 0
\label{212}
\eea
where 
\bea
n_1 \; = \; y \; + \; w & \;\;\;\;\;\;\;\;\;\;\;\;\;\;\; & 
\widetilde{n_1} \; = \; 
- \, y \; + \; w    \label{213} \\
n_2 \; = \; x \; + \; z & \;\;\;\;\;\;\;\;\;\;\;\;\;\;\; & 
\widetilde{n_2} \; = \; 
- \, x \; + \; z    \label{214}
\eea
From (\ref{211}) (for (\ref{212}) it suffices the replacements 
$m_+ \rightarrow m_-$ and $n_1 , n_2 \rightarrow \widetilde{n_1} , 
\widetilde{n_2}$) one gets:
\bea
& & \left( \omega^2 \; - \; k^2 \; - \; m_+^2 \right) \, n_1 \; = \; 0
\label{215} \\
& & n_2 \; = \; \frac{\omega \, + \, \lambda k}{m_+} \, n_1
\label{216}
\eea
so that the dispersion relations (in vacuum) for the $C = \pm 1$ Majorana 
states are
\be
\omega^2 \; - \; k^2 \; = \; m_{\pm}^2
\label{217}
\ee
The time evolution is described by:
\bea
| \, n_{1.2} (t) > & = & e^{-i \, \omega_{1,2} \, t} \; | \, n_{1,2} (0) >
\label{219} \\
| \, \widetilde{n_{1,2}} (t) > & = & e^{-i \, \widetilde{\omega_{1,2}} \, t} \; 
| \,  \widetilde{n_{1,2}} (0) >
\label{220}
\eea
In the ultrarelativistic limit
\bea
\omega_1 & \simeq & k \; + \; \frac{m_+^2}{2k}
\label{221} \\
\widetilde{\omega_1} & \simeq & k \; + \; \frac{m_-^2}{2k}
\label{222}
\eea
From (\ref{219}). (\ref{220}) and (\ref{213}) we may obtain, 
the time evolution of the left-handed weak-interacting neutrino:
\be
| \, y(t) > \; = \; \frac{1}{2} \, \left( e^{-i \, \omega_1 \, t} \; + \;
e^{-i \, \widetilde{\omega_1} \, t} \right) \, | \, y(0) > \; + \; 
\frac{1}{2} \, \left( e^{-i \, \omega_1 \, t} \; - \;
e^{-i \, \widetilde{\omega_1} \, t} \right) \, | \, w(0) >
\label{223}
\ee
If at $t = 0$ we produce (through a 
weak-interaction process) a left-handed neutrino, at later times there 
is a non-vanishing probability of detecting (again through a 
weak-interaction process) the corresponding antiparticle, i.e. a 
right-handed antineutrino. The survival probability of the initial 
$\nu_L$, $P( \nu_L \rightarrow \nu_L ) \, = \, | < y(0) \, | \, y(t) > |^2$, 
is given by
\bea
P( \nu_L \rightarrow \nu_L ) & = & \frac{1}{2} \, \left( 1 \, + \, 
\cos \left( \omega_1 \, - \, \widetilde{\omega_1} \right) t \right) \; = 
\nonumber \\
& \simeq & 1 \; - \, \sin^2 \frac{m_D \, m_M}{4 k} t
\label{224}
\eea
and exhibits a typical oscillatory behaviour. The formula 
(\ref{224}) was first introduced by Pontecorvo \cite{Pontecorvo} in 
analogy to what happens in the $K^0 - \ov{K^0}$ system and, as we have 
shown, can be derived for neutrinos with both Dirac and 
Majorana mass terms. From (\ref{224}) we 
emphasize that both $m_D$ and $m_M$ must be non-vanishing for 
neutrino-antineutrino oscillations to occur.

\section{Neutrino-antineutrino oscillations in matter}

When neutrinos propagate in a medium, their self-energy, acquired 
through the interaction with the particles in the medium, must be 
taken into account in writing the equations of motion; substantially, 
this is done with the replacements of the mass terms with the 
self-energy terms, as explained for example in \cite{Capone}. For 
non-magnetized media (the extension to media with a magnetic field is 
straightforward following \cite{Capone}), at first order in the Fermi 
coupling constant $G_F$, the self-energies 
are given by \cite{Capone}, \cite{Espo2}
\bea
\Sigma^{\prime}_{\nu_L} & = & b_L \, {\not u} \, \frac{1 \, - \, 
\gamma_5}{2}    \label{31}  \\
\Sigma^{\prime}_{\nu_R} & = & 0    \label{32}  \\
\Sigma^{\prime}_{\nu^C_R} & = & - \, b_L \, {\not u} \, 
\frac{1 \, + \, \gamma_5}{2}    \label{33}  \\
\Sigma^{\prime}_{\nu^C_L} & = & 0    \label{34} 
\eea
where $u_{\mu}$ is the medium 4-velocity (we will consider the medium 
rest frame, $u_{\mu} \, = \, (1, \bvec{0})$) and
\be
- \, b_L \; = \; \sqrt{2} \, G_F \, \left( N_e \, - \, \frac{1}{2} \, 
N_n \right)
\label{35}
\ee
for the electron flavour, while
\be
- \, b_L \; = \;  \frac{G_F}{\sqrt{2}} \, N_n 
\label{36}
\ee
for the $\mu$ and $\tau$ flavours, $N_e$ and $N_n$ being the electron 
and neutron number density of the medium respectively. The equations 
of motions are:
\bea
( {\not k} \, - \, m_+ ) \, \nu_R \, + \, ( {\not k} \, - \, m_+ \, + 
\, b_L \, {\not u} ) \, \nu_L & + & 
( {\not k} \, - \, m_+ ) \, \nu^C_L \, + \nonumber \\
& + & ( {\not k} \, - \, m_+ \, 
- \, b_L \, {\not u} ) \, \nu^C_R \; = \; 0
\label{37} \\
- \, ( {\not k} \, - \, m_- ) \, \nu_R \, - \, ( {\not k} \, - \, m_- \, + 
\, b_L \, {\not u} ) \, \nu_L & + & 
( {\not k} \, - \, m_- ) \, \nu^C_L \, + \nonumber \\
& + & ( {\not k} \, - \, m_- \, 
- \, b_L \, {\not u} ) \, \nu^C_R \; = \; 0
\label{38}
\eea
and, explicitely, in the chiral Weyl base,
\bea
& \left. \right. &
\left( \ba{cc}  - \, m_+ & \omega \, + \, b_L \, + \, \bvec{\sigma} 
                \cdot \bvec{k} \\
                \omega \, - \, \bvec{\sigma} \cdot \bvec{k} & - \, m_+
       \ea  \right) \left( 
\ba{c} \nu_R \\  \nu_L \ea \right) 
\; + \nonumber \\
& \left. \right. & + \, \left( 
\ba{cc}  - \, m_+ & \omega \, - \, b_L \, + \, \bvec{\sigma} 
                \cdot \bvec{k} \\
                \omega \, - \, \bvec{\sigma} \cdot \bvec{k} & - \, m_+
       \ea  \right) \, \left( \ba{c} \nu^C_L \\  \nu^C_R \ea \right) 
\; = \; 0   \label{39} \\
& \left. \right. &
\left( \ba{cc}  m_-  & - \, (\omega \, + \, b_L \, + \, \bvec{\sigma} 
                \cdot \bvec{k}) \\
                - \, (\omega \, - \, \bvec{\sigma} \cdot \bvec{k}) & m_-
       \ea  \right)  \left(  
\ba{c} \nu_R \\  \nu_L \ea \right) 
\; + \nonumber \\
& \left. \right. & + \, \left( 
\ba{cc}  - \, m_- & \omega \, - \, b_L \, + \, \bvec{\sigma} 
                \cdot \bvec{k} \\
                \omega \, - \, \bvec{\sigma} \cdot \bvec{k} & - \, m_-
       \ea  \right) \, \left( \ba{c} \nu^C_L \\  \nu^C_R \ea \right) 
\; = \; 0   \label{310} 
\eea
Introducing now, as above, the helicity eigenstates, in the (well 
verified) approximation that the interaction with the medium does not 
change neutrino helicity, the equations (\ref{39}), (\ref{310}) 
take the form
\bea
( \omega \, + \, \lambda \, k ) \, n_1 \; - \; m_+ \, n_2 \; - \, b_L 
\, \widetilde{n_1} & = & 0  \label{311}  \\
- \, m_+ \, n_1 \; + \; ( \omega \, - \, \lambda \, k ) \, n_2 & = & 0 
\label{312}  \\
( \omega \, + \, \lambda \, k ) \, \widetilde{n_1} \; - \; m_- \, 
\widetilde{n_2} \; - \, b_L 
\, n_1 & = & 0  \label{313}  \\
- \, m_- \, \widetilde{n_1} \; + \; ( \omega \, - \, \lambda \, k ) \, 
\widetilde{n_2} & = & 0 
\label{314}
\eea
With simple algebra we can write these equations as 
\bea
\left( \omega^2 \, - \, k^2 \, - \, m_+^2 \right) \, n_2 & = & b_L \, 
\frac{m_+}{m_-} \, \left( \omega \, - \, \lambda \, k \right) \, 
\widetilde{n_2}  \label{315} \\
\left( \omega^2 \, - \, k^2 \, - \, m_-^2 \right) \, \widetilde{n_2} 
& = & b_L \, 
\frac{m_-}{m_+} \, \left( \omega \, - \, \lambda \, k \right) \, 
n_2  \label{316} \\
n_1 & = & \frac{\omega \, - \, \lambda \, k}{m_+} \, n_2
\label{317} \\
\widetilde{n_1} & = & \frac{\omega \, - \, \lambda \, k}{m_-} \, 
\widetilde{n_2}  \label{318} 
\eea
These equations admit non trivial solutions only if
\be
\left( \omega^2 \, - \, k^2 \, - \, m_+^2 \right) \,
\left( \omega^2 \, - \, k^2 \, - \, m_-^2 \right) \; = \; b_L^2
\left( \omega \, - \, \lambda \, k \right)^2
\label{319}
\ee
Eq. ({\ref{319}) is the dispersion relation for Dirac-Majorana 
neutrinos propagating in matter.

Let us now focus on $\lambda = -1$ case; in the ultrarelativistic 
limit, the dispersion relation (\ref{319}) can be viewed as the 
eigenvalue equation for the ``effective hamiltonian''
\be
H_{eff} \; = \; k \; + \; \frac{1}{2 k} \, \left(
\ba{cc} m_+^2 & 0 \\ 0 & m_-^2 \ea \right) \; + \; \left( \ba{cc}
0 & b_L \\ b_L & 0 \ea \right)
\label{320}
\ee
in the base $\left( \ba{c} n_1 \\ \widetilde{n_1} \ea \right)$. 

The time evolution of this states is then obtained by the Sch\"odinger 
equation
\be
i \, \frac{d}{d t} \, \left( \ba{c} n_1 \\ \widetilde{n_1} \ea \right)
\; = \; H_{eff} \, \left( \ba{c} n_1 \\ \widetilde{n_1} \ea \right)
\label{321}
\ee
For the sake of simplicity, we will consider a constant density 
medium; then the hamiltonian in (\ref{321}) can be diagonalized by a 
matrix
\be
U \; = \; \left( \ba{cc} \cos \, \theta_m & \sin \, \theta_m \\
                 - \, \sin \, \theta_m & \cos \, \theta_m
                 \ea  \right)
\label{322}
\ee
(neglecting CP violating effects) depending on a constant mixing angle 
between $n_1$ and $\widetilde{n_1}$. The eigenstates are then
\be
\left( \ba{c} n_{1m} \\ \widetilde{n_{1m}} \ea \right) \; = \; 
\left( \ba{cc} \cos \, \theta_m & - \, \sin \, \theta_m \\
                 \sin \, \theta_m & \cos \, \theta_m
                 \ea  \right)
\left( \ba{c} n_1 \\ \widetilde{n_1} \ea \right)
\label{323}
\ee
and the matrix $U^T \, H_{eff} \, U$ becomes diagonal when the mixing 
angle is given by
\be
\tan \, 2 \theta_m \; = \; - \, \frac{4 \, b_L \, k}{m_D \, m_M}
\label{324}
\ee
The time evolution of the physical matter eigenstates is
\bea
| \, n_{1m} (t) > & = & e^{-i \, \omega_- \, t} \; | \, n_{1m} (0) >
\label{325} \\
| \, \widetilde{n_{1m}} (t) > & = & e^{-i \, \widetilde{\omega_+} \, t} \; 
| \,  \widetilde{n_{1m}} (0) >
\label{326}
\eea
and the eigenvalues are:
\be
\omega_{\pm} \; = \; k \; + \; \frac{m_D^2 \, + \, m_M^2}{8 k} \; \pm 
\; \frac{1}{2} \, \sqrt{\left( \frac{m_D \, m_M}{2 k} \right)^2 \; + 
\; 4 b_L^2}
\label{327}
\ee
From the time dependence of the weak-interaction eigenstate 
$\nu_L$,
\bea
| \, y(t) > & = & \frac{1}{2} \left( \left( e^{-i \, \omega_- \, t} \; 
+ \; e^{-i \, \omega_+ \, t} \right) \; + \; \sin \, 2 \theta_m \, 
\left( e^{-i \, \omega_- \, t} \; - \; e^{-i \, \omega_+ \, t} \right) 
\right) \, | \, y(0) > \; +  \nonumber \\
& + & \frac{1}{2} \, \cos \, 2 \theta_m \, 
\left( e^{-i \, \omega_- \, t} \; - \; e^{-i \, \omega_+ \, t} \right) 
\, | \, w(0) >
\label{328}
\eea
we can finally obtain the expression for the survival probability for 
an initial produced $\nu_L$:
\bea
P(\nu_L \rightarrow \nu_L) & = & 1 \; - \; \cos^2 \, 2 \theta_m \, 
\sin^2 \, \frac{\omega_+ \, - \, \omega_-}{2} \, t  \; =  \nonumber \\
& \simeq & 1 \; - \; \left( 1 \, + \, \left( \frac{4 \, b_L \, k}{m_D 
\, m_M} \right)^2 \right)^{- 1} \, \sin^2 \, \left( \sqrt{\left( 
\frac{m_D \, m_M}{2 k} \right)^2 \; + \; 4 \, b_L^2} \; \frac{t}{2} 
\right)
\label{329}
\eea
We immediately note, from (\ref{329}), a completely different 
behaviour of neutrino-antineutrino oscillations in matter with respect 
to the case of flavour oscillations. In fact, matter 
neutrino-antineutrino oscillations have both an amplitude and a period 
(and then an oscillation lenght) {\it smaller} than those 
corresponding to vacuum oscillations. In other words, opposite
to what happens for flavour oscillations, 
neutrino-antineutrino oscillations in matter are {\it always} 
suppressed with respect to the vacuum case; furthermore, no resonance 
condition (analogous to the MSW effect \cite{MSW}) can occur.

\section{Limits on the Majorana neutrino mass}

The fact that neutrino-antineutrino oscillations in matter are suppressed 
with respect to the vacuum  makes this phenomenon pratically 
non-interesting for astrophysical scopes. In particular, there cannot 
be a matter solutions in terms of neutrino-antineutrino oscillations 
to the solar neutrino problem \cite{SNP}. This can be easily seen from 
the following arguments. The amplitude $cos^{2}2\theta_{m}$ in (\ref{329}) 
is lower for more energetic neutrinos (such as those produced by the 
$Be^{7}$ and $B^{8}$ reactions in the sun \cite{Bahcall}) than for less
energetic ones (such as those from the p-p chain). So one predicts a 
smaller reduction for $Be^{7}$ and $B^{8}$ neutrinos than for
p-p neutrinos.
This is just the opposite of what is needed to explain the 
experimental results \cite{opposite}, so 
the matter effects on neutrino-antineutrino oscillations cannot help 
to solve the solar neutrino problem.

On the contrary, the analysis of the experimental data on solar 
neutrino \cite{SNP} shows that a vacuum neutrino-=antineutrino 
oscillations solution is indeed possible with the following value of 
the product of Dirac and Majorana $\nu_{e}$ masses:
\be
m_{D} \; m_{M} \; \simeq 6 \times 10^{-11} \; eV^{2}
\label{41}
\ee
Note that this solution is equivalent to the vacuum  
flavour oscillation solution 
with maximal (flavour) mixing angle ($\sin^2 \, 
2 \theta \, \simeq \, 1$) and $\Delta m^2 \, = \, m^2_{\nu_2} \, - \, 
m^2_{\nu_1} \, \simeq \, 6 \, \times \, 10^{- 11} \; eV^2$ 
\cite{Lisi}.

However, the best probe for neutrino-antineutrino oscillations (in 
vacuum) is with laboratory experiments.

We have analyzed the current data on disappearance laboratory 
experiments for giving limits on $m_D \, m_M$. The most stringent 
limit for the electron flavour comes from the experiment at the 
Krasnoyarsk reactor \cite{Kras}; with the formula (\ref{224}) for the 
survival probability we have obtained
\be
m_D^{\nu_e} \, m_M^{\nu_e} \; \leq \; 7.5 \, \times \, 10^{-3} \; eV^2
\;\;\;\;\;\;\;\;\;\;\;\;\;\;\; (90 \% \; C.L.)
\label{42}
\ee
For the muon flavour, experiments allow both low and large values for 
$m_D^{\nu_{\mu}} \, m_M^{\nu_{\mu}}$ (see \cite{PDG}); from the 
Particle Data Group analysis \cite{PDG} we deduce
\be
m_D^{\nu_{\mu}} \, m_M^{\nu_{\mu}} \; \leq \; 0.23 \; eV^2
\label{43}
\ee
or 
\be
m_D^{\nu_{\mu}} \, m_M^{\nu_{\mu}} \; \geq \; 1500 \; eV^2
\label{44}
\ee
Given the form (\ref{224}) for the survival probability, from the 
experiments one can only deduce limits on the product of the Dirac 
times the Majorana mass. However, if the neutrino Dirac mass terms are 
generated by the same Higgs doublet giving mass to the charged fermion 
(as happens in G.U.T.), it is natural to assume that $m_D$ is of the 
same order of magnitude of the (Dirac) mass of the charged fermion 
associated to the given neutrino in the same supemultiplet (for 
example, in SO(10) $\nu_e$, $\nu_{\mu}$, $\nu_{\tau}$ are linked to 
the up-quarks $u,c,t$). Assuming then $m_D^{\nu_e} \approx O(MeV)$ and 
$m_D^{\nu_{\mu}} \approx O(100 \div 1000 \, MeV)$, from (\ref{43}), 
(\ref{44}) one can deduce the following indicative limits on $\nu_e$ 
and $\nu_{\mu}$ Majorana masses:
\be
m_M^{\nu_{e}} \; \leq \; 10^{-8} \; eV
\ee
and
\be
m_M^{\nu_{\mu}} \; \leq \; 10^{-9} \; eV
\ee
or
\be
m_M^{\nu_{\mu}} \; \geq \; 10^{-6} \; eV
\ee
Note that the limit for $m_M^{\nu_{e}}$ is smaller by
several orders of magnitude than the ones 
obtained from neutrinoless double $\beta$ decay (some $eV$) \cite{double}.

\section{Conclusions}

In this paper we have analyzed the theory of Pontecorvo 
neutrino-antineutrino oscillations in vacuum and in matter, showing 
that they can occur only if neutrinos are Majorana particles with 
both Dirac and Majorana masses non-vanishing. With a field-theoretic 
approach we have rederived the Pontecorvo formula (eq. (\ref{224})) for 
vacuum oscillations and then generalized this formula to matter 
oscillations, finding that neutrino-antineutrino transitions in matter 
are suppressed with respect to vacuum ones. 

Furthermore, we have shown 
that for solar neutrinos, the matter effects
are not in agreement with the experimental data, so that 
matter Pontecorvo oscillations does not explain the solar neutrino 
problem. However, a vacuum solution to this problem in terms of 
neutrino-antineutrino transitions indeed exists, and we give the 
corresponding range of the product of $\nu_e$ Dirac and Majorana 
mass. 

The existing limits from laboratory experiments for this product 
for electron and muon flavours have also been analyzed, leading to 
limits on $\nu_e$ Majorana mass (in the framework of Grand Unified Theories) 
several orders of magnitude smaller than the one found from the 
experiments on neutrinoless double beta decay.

\vspace{1truecm}

\noindent
{\Large \bf Acknowledgements}\\
\noindent
The authors express their sincere thanks to Prof. F. Buccella.

\end{document}